\documentclass[12pt]{article}

\usepackage{amssymb}
\usepackage[mathscr]{euscript}
\usepackage{graphicx}
\usepackage[all]{xypic}

\newtheorem{lemm}{Lemma}[section]
\newtheorem{theo}{Theorem}[section]

\newcommand{\iN}{\hbox{ {\leaders\hrule\hskip.2cm}{\vrule height .22cm} }}

\newcommand{\R}{\Bbb{R}}
\newcommand{\C}{\Bbb{C}}

\newcommand{\Oct}{\Bbb{O}}
\renewcommand{\H}{\Bbb{H}}

\title{Some examples of `second order elliptic integrable systems associated to a 4-symmetric space'}
\author{Fr{\'e}d{\'e}ric {\sc H{\'e}lein}, Universit{\'e} Paris 7}
\date{Saturday 11 August 2006}
\begin{document}
\maketitle
\begin{center}
{\em London Mathematical Society} Durham symposium\\ {\sc Methods of Integrable Systems in Geometry}\\ 11--21 August 2006
\end{center}

\section{Hamiltonian Stationary Lagrangian (HSL) surfaces}
\subsection{A variational problem in $\R^4$}
$\R^4$ has the canonical Euclidean structure $\langle\cdot,\cdot\rangle$ and the symplectic form $\omega:= dx^1\wedge dx^2 + dx^3\wedge dx^4$. An immersed surface $\Sigma\subset \R^4$ is
\begin{enumerate}
\item {\bf Lagrangian} iff $\omega_{|\Sigma} = 0$
\item {\bf Hamiltonian Stationary Lagrangian} (HSL) iff $\omega_{|\Sigma} = 0$ and $\Sigma$ is a critical point of the area functional ${\cal A}$ with respect to all {\em Hamiltonian vector fields} $\xi_h$ s.t.:
\begin{itemize}
\item  $\exists h \in {\cal C}^\infty_c(\R^4)$, $\xi_h\iN \omega+dh = 0$ \item equivalentely, if $J$ is the complex structure s.t. $\omega = \langle J\cdot,\cdot\rangle$, $\xi_h = J\nabla h$.
\end{itemize}
It means that $\delta{\cal A}_\Sigma(\xi_h) = 0$, $\forall h \in {\cal C}^\infty_c(\R^4)$.
\end{enumerate}
What is the Euler equation ?\\
The Gauss map is:
\[
\begin{array}{ccccc}
\gamma:  \Sigma & \longrightarrow & Gr_{Lag}(\R^4) & \subset & Gr_2(\R^4)\\
& & S^1\times S^2 & \subset & S^2\times S^2
\end{array}
\]
Denote $\gamma = (\rho_\Sigma,\sigma_\Sigma)$ the two components of $\gamma$. For a Lagrangian immersion $\rho_\Sigma\simeq e^{i\beta}$. Then the {\em mean curvature vector} is
\[
\vec{H} = J\nabla \beta.
\]
\begin{lemm} $\Sigma$ is HSL iff
\[
\left\{\begin{array}{c} \omega_{|\Sigma} = 0\\ \Delta_{\Sigma}\beta = 0.
\end{array}\right.
\]
\end{lemm}
{\em Remark:} $\Sigma$ is special Lagrangian iff $\displaystyle \left\{\begin{array}{c} \omega_{|\Sigma} = 0\\ \beta = \hbox{Constant}.
\end{array}\right. \quad \Longleftrightarrow \quad \left\{\begin{array}{c} \omega_{|\Sigma} = 0\\ \Sigma\hbox{ is minimal}
\end{array}\right.$.\\

\noindent
An analytic study was done by R. {\sc Schoen} and J. {\sc Wolfson} \cite{sw} (in a 4-dimensional Calabi--Yau manifold).

\subsection{It is a completely integrable system (F.H.--P. {\sc Romon} \cite{hr1,hr2})}
Let $\Omega\subset \C$ be an open subset and $X:\Omega\longrightarrow \R^4$ a (local) conformal parametrization of $\Sigma$. Set
\[
\rho_X:= \rho_\Sigma\circ X,
\]
the {\em left Gauss map}.\\

\noindent
{\bf Idea:} to lift the pair $(X,\rho_X)$ to a map $F:\Omega\longrightarrow \mathfrak{G}$, where $\mathfrak{G}$ is a local symmetry group of the problem. The more naive choice is $\mathfrak{G} = SO(4)\ltimes \R^4$, the group of isometries of $\R^4$. Then
\[
F = \left( \begin{array}{cc}R & X\\0 & 1\end{array}\right)\simeq (R,X),
\]
where $R:\Omega\longrightarrow SO(4)$ encodes $\rho_X \simeq e^{i\beta}$. (Alternatively one can choose $\mathfrak{G} = U(2)\ltimes \C^2$, with the identification $\C^2\simeq (\R^4,J)$ and $U(2)$: subgroup of $SO(4)$. Then the way $R\in U(2)$ encodes $\beta$ is simply through the relation $\hbox{det}_\C R = e^{i\beta}$).\\

\noindent
In all cases there exists an automorphism $\tau:\mathfrak{G}\longrightarrow \mathfrak{G}$ s.t. $\tau^4 = Id$. This automorphism acts on the Lie algebra $\mathfrak{g}$ and can be diagonalized with the eigenvalues $i$, $1$, $-i$ and $-1$. Hence the vector space decomposition
\[
\begin{array}{cccccccc}
\mathfrak{g}^\C = & \mathfrak{g}_{-1} & \oplus & \mathfrak{g}_0^\C & \oplus & \mathfrak{g}_1 & \oplus & \mathfrak{g}_2^\C \\
\hbox{eigenvalues: }& -i && 1 && i && -1
\end{array}
\]
Then consider the (pull-back of the) Maurer--Cartan form
\[
\alpha:= F^{-1}dF
\]
and split $\alpha = \alpha_{-1} + \alpha_0+ \alpha_1 + \alpha_2$ according to this decomposition. Then do the further splitting $\alpha_2 = \alpha_2' + \alpha_2''$, where $\alpha_2' = \alpha({\partial \over \partial z}) dz$ and $\alpha_2''= \overline{\alpha_2'}$. And consider the family of deformations
\[
\alpha_\lambda:= \lambda^{-2} \alpha_2' + \lambda^{-1}\alpha_{-1} + \alpha_0 + \lambda\alpha_1 + \lambda^2\alpha_2'',
\quad \lambda\in \C^*.
\]
Then:\\

\noindent
\begin{theo} \begin{enumerate}
\item $X$ is Lagrangian iff $\alpha_{-1}'' = 0$
\item $X$ is HSL iff $\alpha_{-1}'' = 0$ and, $\forall \lambda\in \C^*$, $d\alpha_\lambda + {1\over 2} [\alpha_\lambda\wedge \alpha_\lambda] = 0$.
\end{enumerate}
\end{theo}
Using this characterisation one see easily that HSL surfaces are solutions of a completely integrable system.\\

\noindent
Note that analogous formulations work for HSL surfaces in $\C P^2 = SU(3)/S(U(2)\times U(1)) = U(3)/U(2)\times U(1)$, $\C D^2 = SU(2,1)/S(U(2)\times U(1))$, $\C P^1\times \C P^1$, $\C D^1\times \C D^1$ \cite{hr3}.

\section{Generalizations in $\R^4$ (after I. {\sc Khemar} \cite{k1})}
Again $\R^4$ is endowed with its canonical Euclidean structure. We will also use an identification of $\R^4$ with the quaternions $\H$.
%\footnote{note that this rests on the choice of a vector space embedding $r:\R\longrightarrow \H$, where $\R\simeq r(\R)$ is the line of real quaternions and an orientation of $\hbox{Im}\H:= \R^\perp$, which provides us with a vector product $\times$ in $\hbox{Im}\H$, then the quaternionic structure is defined by $\forall (t,u),(s,v)\in \R\oplus \hbox{Im}\H$, $(t+u)(s+v) = (ts -\langle u,v\rangle) + (tv + su + u\times v)$.}. 
We recall that this allows to represent rotations $R\in SO(4)$ by a pair $(p,q)\in S^3\times S^3\subset \H\times \H$ of unit quaternions such that $\forall z\in \H$, $R(z) = pz\overline{q}$. In other words, denoting by $L_p:z\longmapsto pz$ and $R_{\overline{q}}:z\longmapsto z\overline{q}$, we have $R = L_pR_{\overline{q}} = R_{\overline{q}}L_p$. The pair $(p,q)$ is unique up to sign, hence the identification $SO(4)\simeq S^3\times S^3/\{\pm \}$.\\

\noindent
Moreover we can also precise the identification $Gr_2(\R^4) \simeq S^2\times S^2$. Let 
\[
Stiefel_2(\H):= \{ (e_1,e_2)\in \H\times\H |\, |e_1| = |e_2| = 1, \langle e_1,e_2\rangle = 0\}.
\]
Observe that $\forall (e_1,e_2)\in Stiefel_2(\H)$, $e_2\overline{e_1}$ (resp. $\overline{e_1}e_2$) is unitary (because $e_1$ and $e_2$ are so) and imaginary (because $\langle e_1,e_2\rangle = 0$). Hence this defines two maps
\[
\begin{array}{ccc} Stiefel_2(\H) & \longrightarrow & S^2 \\
(e_1,e_2) & \longmapsto & e_2\overline{e_1}
\end{array},\quad
\begin{array}{ccc} Stiefel_2(\H) & \longrightarrow & S^2 \\
(e_1,e_2) & \longmapsto & \overline{e_1}e_2
\end{array}.
\]
These maps factor through the natural map $P:(e_1,e_2)\longmapsto \hbox{Span}\{e_1,e_2\}$
from $Stiefel_2(\H)$ to the oriented Grassmannian $Gr_2(\H)$: let
\[
\begin{array}{c}\rho:Gr_2(\H)\longrightarrow S^2\\ \hbox{s. t. }\rho\circ P(e_1,e_2) = e_2\overline{e_1}
\end{array},\quad
\begin{array}{c}\sigma:Gr_2(\H)\longrightarrow S^2\\ \hbox{s. t. }\sigma\circ P(e_1,e_2) = \overline{e_1}e_2.
\end{array}
\]
Then $(\rho,\sigma):Gr_2(\H)\longrightarrow S^2\times S^2$ is a diffeomorphism.\\

\subsection{Immersions of a surface in $\H$ with a harmonic `left Gauss map'}

Let $X:\Omega\longrightarrow \H$ be a conformal immersion and $\rho_X:\Omega\longrightarrow S^2$ its {\em left Gauss map}, i.e. $\forall z\in \Omega$, $\rho_X(z)$ is the image of $\hbox{Span}({\partial X\over \partial x}(z), {\partial X\over \partial y}(z))$ by $\rho$. It is characterised by
\[
{\partial X\over \partial y} = \rho_X{\partial X\over \partial x}
\quad \Longleftrightarrow\quad
i{\partial X\over \partial z} = \rho_X{\partial X\over \partial z}.
\]
(In the second equation the $i$ on the l.h.s. is the complex structure on $\Omega\subset \C$, whereas the $\rho_X$ on the r.h.s. denotes the left multiplication in $\H$.)\\

\noindent
{\bf Remark:} instead of viewing $\rho_X$ as the left component of the Gauss map in $Gr_2(\H)\simeq S^2\times S^2$, an alternative interpretation is that $\rho_X$ is a map into the `left' connected component of the manifold of compatible complex structures ${\cal J}_{\H}\simeq S^2\cup S^2$ on $\H$ (cf. the work of F. {\sc Burstall}). \\

\noindent
{\bf Idea:} to lift the pair $(X,\rho_X)$ by a framing $F:\Omega\longrightarrow \mathfrak{G}$, $\mathfrak{G}$ is a subgroup of $SO(4)\ltimes \R^4$.

\noindent
{\bf How ?} We fix some constant imaginary unit vector $u\in S^2\subset \hbox{Im}\H$.\\
\begin{itemize}
\item {\em First method:} we lift $X$ {\bf and} its full Gauss map $T_X\Sigma\simeq (\rho_X,\sigma_X)$: we let $(e_1,e_2)$ be any moving frame which is an orthonormal basis of $T_{X(z)}\Sigma$ (e.g. $e_1 = {\partial X\over \partial x}/|{\partial X\over \partial x}|$, $e_2 = {\partial X\over \partial x}/|{\partial X\over \partial y}|$) and we choose $F = (R,X)$ s.t. $R$ satisfies:
\[
R(1) = e_1,\quad R(u) = e_2.
\]
Decompose $R = L_pR_{\overline{q}}$, then
\[
R(1) = p\overline{q}, R(u) = pu\overline{q}, \hbox{ so that} \quad
\rho_X = e_2\overline{e_1} = pu\overline{p}.
\]
{\em Note:} In this case we must choose $\mathfrak{G} = SO(4)\ltimes \R^4$ (which acts transitively on $Stiefel_2(\H)$).

\item
{\em Second method:} we lift {\bf only} $X$ and $\rho_X$. Then it means that we choose $F = (R,X)$, where $R = L_pR_{\overline{q}}$ is s.t.
\[
\rho_X = pu\overline{p}.
\]
Hence the choice of $q$ is not relevant. In other words introducing the {\em (left) Hopf fibration}
\[
\begin{array}{cccc}
{\cal H}^u_L: & SO(4) & \longrightarrow & S^2\\
& L_pR_{\overline{q}} & \longmapsto & pu\overline{p},
\end{array}
\]
we choose the lift $F = (R,X)$ in such a way that ${\cal H}^u_L\circ R = \rho_X$.\\
We observe that in this case one may choose $q = 1$ and assume that $R\in \{L_p|\, p\in S^3\}\simeq Spin3$, i.e. work with $\mathfrak{G} = Spin3\ltimes \H$. The restriction of ${\cal H}^u_L$ to $Spin3$ (viewed as a subgroup of $SO(4)$) is just the Hopf fibration ${\cal H}^u: S^3\longrightarrow S^2$.
\end{itemize}
Actually the second point of view is more general and leads to a simpler theory.\\

\noindent
Now let $\tau: (R,X)\longmapsto (L_uRL_u^{-1},-L_uX)$, a 4th order automorphism of $\mathfrak{G}$ (i.e. $\tau^4 = Id$). It induces a 4th order automorphism on its Lie algebra $\mathfrak{g}$. Let
\[
\mathfrak{g}^\C = \mathfrak{g}_{-1}\oplus \mathfrak{g}_0^\C \oplus \mathfrak{g}_1 \oplus \mathfrak{g}_2^\C
\]
be its associated eigenspace decomposition. Split the Maurer--Cartan form $\alpha = F^{-1}dF$ according to this decomposition: $\alpha = \alpha_{-1} + \alpha_0 + \alpha_1 + \alpha_2$ and let
\[
\beta_{\lambda^2}:= \lambda^{-2}\alpha_2' + \alpha_0 + \lambda^2\alpha_2'',
\]
\[
\alpha_\lambda:= \lambda^{-2} \alpha_2' + \lambda^{-1}\alpha_{-1} + \alpha_0 + \lambda\alpha_1 + \lambda^2\alpha_2'' = \beta_{\lambda^2} + \lambda^{-1}\alpha_{-1} + \lambda\alpha_1.
\]
Then:\\

\noindent
\begin{lemm}\label{Lemma1} If $X:\Omega\longrightarrow \R^4$ is a conformal immersion and if $R:\Omega\longrightarrow SO(4)$ is an arbitrary smooth map, then
\[
{\cal H}^u_L\circ R = \rho_X\quad
\Longleftrightarrow \quad
\alpha_{-1}'' = 0.
\]
In other words $F=(R,X):\Omega\longrightarrow SO(4)\ltimes\R^4$ lifts $(X,\rho_X)$ iff $\alpha_{-1}'' = 0$.
\end{lemm}
{\em Remark:} $\alpha_1$ is the complex conjugate of $\alpha_{-1}$, so that $\alpha_{-1}'' = 0$ iff $\alpha_1' = 0$.\\

\noindent
\begin{lemm}\label{Lemma2} We have:
\begin{equation}\label{identity}
d\alpha_\lambda + {1\over 2}[\alpha_\lambda\wedge \alpha_\lambda] = d\beta_{\lambda^2} + {1\over 2}[\beta_{\lambda^2}\wedge \beta_{\lambda^2}] + (\lambda^{-3}-\lambda) [\alpha_2'\wedge \alpha_{-1}''] + (\lambda^3-\lambda^{-1}) [\alpha_2''\wedge \alpha_1'].
\end{equation}
Hence in particular, if $F$ lifts $(X,\rho_X)$, then $d\alpha_\lambda + {1\over 2}[\alpha_\lambda\wedge \alpha_\lambda] = d\beta_{\lambda^2} + {1\over 2}[\beta_{\lambda^2}\wedge \beta_{\lambda^2}]$.
\end{lemm}

\noindent
In order to interpret (\ref{identity}) we further observe that
\begin{enumerate}
\item $\mathfrak{G}^\tau$, the fixed subset of $\tau:\mathfrak{G}\longrightarrow \mathfrak{G}$, is a subgroup of $\mathfrak{G}$ with Lie algebra $\mathfrak{g}_0$
\item $\mathfrak{G}^{\tau^2} = \{(R,0)\in \mathfrak{G}\}$, the fixed subset of $\tau^2:\mathfrak{G}\longrightarrow \mathfrak{G}$, is a subgroup of $\mathfrak{G}$ with Lie algebra $\mathfrak{g}_0\oplus \mathfrak{g}_2$,
\end{enumerate}
with the inclusions
\[
\mathfrak{G}^\tau\subset \mathfrak{G}^{\tau^2} \subset \mathfrak{G}.
\]
Moreover $\mathfrak{G}/\mathfrak{G}^{\tau^2} \simeq \H$ and $\mathfrak{G}^{\tau^2}/\mathfrak{G}^\tau\simeq S^2$ and the projection map
\[
\begin{array}{ccc}
\mathfrak{G}^{\tau^2} & \longrightarrow & \mathfrak{G}^{\tau^2}/\mathfrak{G}^\tau \simeq S^2 \\
R\simeq (R,0) & \longmapsto & R\hbox{ mod }\mathfrak{G}^\tau
\end{array}
\]
coincides with the Hopf fibration ${\cal H}^u_L$. Hence, by applying the standard theory of harmonic maps into symmetric spaces, we deduce that:
\[
d\beta_{\lambda^2} + {1\over 2}[\beta_{\lambda^2}\wedge \beta_{\lambda^2}] = 0
\quad \Longleftrightarrow \quad
{\cal H}^u_L\circ R:\Omega\longrightarrow S^2\hbox{ is harmonic}.
\]
Putting Lemmas \ref{Lemma1} and \ref{Lemma2} and these observations together we conclude with the following:\\

\begin{theo} Let $X:\Omega\longrightarrow \H$ be a conformal immersion and $\rho_X:\Omega\longrightarrow S^2$ its left Gauss map. Let $F = (R,X):\Omega\longrightarrow \mathfrak{G}$ be any smooth map. Then
\begin{enumerate}
\item ${\cal H}^u_L\circ R = \rho_X$ (i.e. $F$ is a lift of $(X,\rho_X)$) iff $\alpha_{-1}'' = 0$
\item If so, i.e. if $F$ is a lift of $(X,\rho_X)$, then $\rho_X$ is harmonic iff
\[
d\alpha_\lambda + {1\over 2}[\alpha_\lambda\wedge \alpha_\lambda] = 0.
\]
\end{enumerate}
\end{theo}
\subsection{Examples}
\subsubsection{HSL surfaces revisited}
Let us introduce again the symplectic form $\omega = dx^1\wedge dx^2 + dx^3\wedge dx^4$. Note that $\omega = \omega_1:= \langle L_i\cdot,\cdot \rangle$. Let us introduce also $\omega_2:= \langle L_j\cdot,\cdot \rangle = dx^1\wedge dx^3 + dx^4\wedge dx^2$ and $\omega_3:= \langle L_k\cdot,\cdot \rangle = dx^1\wedge dx^4 + dx^2\wedge dx^4$. Then
\[
e_2\overline{e_1} = \rho(e_1,e_2) = i\omega_1(e_1,e_2) + j\omega_2(e_1,e_2) + k\omega_3(e_1,e_2).
\]
So $X$ is a conformal {\em Lagrangian} immersion iff $X^*\omega_1 =0$, i.e. iff $\rho_X$ takes values in
\[
S^1 = \{j\cos\beta + k\sin\beta = e^{i\beta}j|\, \beta\in \R\}.
\]
Hence a lift of $(X,\rho_X)$ is characterized by
\[
pu\overline{p} = {\cal H}^u_L\circ R = \rho_X = e^{i\beta}j.
\]
A convenient choice for $u$ is to assume that $u\perp i$, e.g. $u = j$. In that case
\[
\{p\in S^3|\, pu\overline{p} = e^{i\beta}j\} = \{e^{i\beta/2}e^{j\theta}|\, \theta\in \R\}
\]
and the simplest choices are $p = \pm e^{i\beta/2}$.\\

\noindent
With this choice:
\begin{itemize}
\item if we start with the group $\mathfrak{G} = SO(4)\ltimes \R^4$, our lift satisfies $R = L_{e^{i\beta/2}}R_{\overline{q}}$, i.e. we can reduce $SO(4)\ltimes \R^4$ to $U(2)\ltimes\C^2$
\item if we start with the group $\mathfrak{G} = Spin3\ltimes \H$, our lift satisfies $R = L_{e^{i\beta/2}}$, i.e. we can reduce $Spin3\ltimes \R^4$ to $U(1)\ltimes\C^2$ (cf. spinor lifts, related to the {\sc Konopelchenko}--{\sc Taimanov} representation formula).
\end{itemize}

\subsubsection{Constant mean curvature surfaces in $\R^3$}
Consider an immersed surface $\Sigma$ in $\H$ with a harmonic left Gauss map. If we assume further that this surface is contained in $\hbox{Im}\H$, then any orthonormal basis $(e_1,e_2)$ of $T_{X(z)}\Sigma$ is composed of imaginary vectors. Hence
\[
\rho_X = e_2\overline{e_1} = - \overline{e_1}e_2 = - \sigma_X,
\]
so that $\rho_X$ is harmonic iff $\sigma_X$ is so. Actually $\rho_X$ is nothing but the Gauss map of $\Sigma$ in $\hbox{Im}\H\simeq \R^3$. Hence by Ruh--Vilms theorem we know that $\Sigma$ is a {\em constant mean curvature surface} in $\R^3$. Conversely any constant mean curvature surface in $\R^3$ arises that way.

\subsection{Other generalizations in dimension 4}
This theory can be generalized to surfaces in $S^4$ or $\C P^2$: then $(X,\rho_X)$ is replaced by a lift of the immersion $X$ in the four dimensional manifold into the twistor bundle of complex structures. The condition of $\rho_X$ being harmonic is replaced by the fact this lift is vertically harmonic (the fiber being the set of (left) compatible complex structures, diffeomorphic to $S^2$). This follows from independant works by F. {\sc Burstall} and I. {\sc Khemar}.

\section{A generalization for surfaces in $\R^8$ (I. {\sc Khemar} \cite{k1})}
The following theory is based on the identification of $\R^8$ with octonions $\Oct$. Again the map
\[
\begin{array}{ccc}
Stiefel_2(\Oct) & \longrightarrow & S^6\\
(e_1,e_2) & \longmapsto & e_2\overline{e_1},
\end{array}
\]
where $S^6\in \hbox{Im}\Oct\subset \Oct$, can be factorized through the map $P: Stiefel_2(\Oct) \longrightarrow Gr_2(\Oct)$, $(e_1,e_2)\longmapsto \hbox{Span}\{e_1,e_2\}$ by introducing
\[
\begin{array}{c}
\rho:  Gr_2(\Oct) \longrightarrow S^6\\
\hbox{s.t. } \rho\circ P(e_1,e_2) = e_2\overline{e_1}.
\end{array}
\]
Let $\Sigma$ be an immersed surface in $\Oct$ we say that {\em $\Sigma$ is $\rho$-harmonic iff the composition of the Gauss map $\Sigma\longrightarrow Gr_2(\Oct)$ with $\rho$ is harmonic}.\\

\noindent
This theory is completely similar with the theory of surfaces in quaternions $\H$ which used the group $\mathfrak{G} = Spin3\ltimes \H$, where $Spin3$ can be seen as the subgroup of $SO(4)$ generated by $L_i$, $L_j$ and $L_k$ and the induced representation of $Spin3$ was the spinor representation $\H$. Here we will use $\mathfrak{G} = Spin7\ltimes \Oct$, where $Spin7$ can be identified with the subgroup of $SO(8)$ generated by $\{L_v|\, v\in S^6\subset\hbox{Im}\Oct\}$ and the induced representation on $\R^8$ coincides with the spinor representation of $Spin 7$ on $\Oct$. A difference however is that $Spin7$ is "bigger" than $Spin3$ and in particular acts transitively on $Stiefel_2(\Oct)$ (with isotropy $SU(3)$) and $Gr_2(\Oct)$ (with isotropy $G_2$), whereas $Spin3$ do not act transitively on $Gr_2(\H)$. After fixing an imaginary unit octonion $u\in \Oct$, a `Hopf' fibration
\[
\begin{array}{cccl}
{\cal H}^u:& Spin7 & \longrightarrow & S^6\\
& p & \longmapsto & {\cal H}^u(p),\hbox{ s.t. }pL_up^{-1} = L_{{\cal H}^u(p)}
\end{array}
\]
can be defined.\\

\noindent
Now let $X:\C\supset \Omega \longrightarrow \Oct$ be a conformal immersion and denote $\rho_X:= \rho\circ T_X\Sigma$ the composition of the Gauss map $T_X\Sigma$ of $X$ with $\rho$. After having fixed $u\in S^6\subset \Oct$ we let
\[
F = \left(\begin{array}{cc}R & X\\0 & 1\end{array}\right) \simeq (R,X):\Omega \longrightarrow Spin7\ltimes \Oct,
\]
be a smooth map. We say that $F$ lifts $(X,\rho_X)$ iff ${\cal H}^u\circ R = \rho_X$. Using the 4th order automorphism $\tau:\mathfrak{G}\longrightarrow \mathfrak{G}$ defined by
\[
\tau (R,X) = (L_uRL_u^{-1},-L_uX),
\]
we can characterized among all maps $F = (R,X)$ those which lift $\rho_X$ by the condition $\alpha_{-1}'' = 0$ (after a decomposition of the Maurer--Cartan form $\alpha:= F^{-1}dF$ along the eigenspaces of the action of $\tau$ on the Lie algebra $\mathfrak{g}$ of $\mathfrak{G}$). Then the $\rho$-harmonic immersions satisfy a zero curvature equation $d\alpha_\lambda + {1\over 2}[\alpha_\lambda\wedge \alpha_\lambda] = 0$ similar to the previous case.\\

\noindent
Again $\rho_X$ can be interpreted as a map into the manifold ${\cal J}_{\Oct}$ of compatible complex structures on $\Oct$, because of the relation $\rho_X{\partial X\over \partial z} = i{\partial X\over \partial z}$. However the embedding $S^6\subset {\cal J}_{\Oct}$ is much less clear than the inclusion $S^2\subset {\cal J}_{\H}$ that we used previously: we recall indeed that ${\cal J}_{\H}\simeq S^2_L\cup S^2_R$ and hence that our $S^2$ was just the (left) connected component of ${\cal J}_{\H}$. However ${\cal J}_{\Oct} \simeq SO(8)/U(4)$ is 12 dimensional, so that our $S^6$ is now a particular submanifold of ${\cal J}_{\Oct}$. Hence a twistor intepretation of the theory in $\Oct$ seems less clear.

\section{Towards a supersymmetric interpretation}
{\bf Observation :} the coefficients of $\alpha_{-1}$ and $\alpha_1$ actually behave like spinors (they turn half less than those of  $\alpha_2$ when $\lambda$ run over $S^1$ and they satisfy a kind of Dirac equation). This motivates the following results by I. {\sc Khemar} \cite{k2}.

\subsection{Superharmonic maps into a symmetric space}
For simplicity we restrict ourself to maps into the sphere $S^n\subset \R^{n+1}$. It can be seen as a system of PDE's on a map $u:\Omega\longrightarrow S^n$ (where $\Omega\subset \C$) and {\em odd} sections $\psi_1$, $\psi_2$ of $u^*TS^n$. This system is
\begin{equation}\label{superharmonic2+1}
 \left\{ \begin{array}{ccl}
          \displaystyle \nabla_{\overline{z}}{\partial u\over \partial z} & = & \displaystyle {1\over 4}\left(\psi \langle \psi, {\partial u\over \partial \overline{z}} \rangle - \overline{\psi}\langle \overline{\psi},{\partial u\over \partial z} \rangle\right)\\
\displaystyle \nabla_{\overline{z}}\psi & = & \displaystyle {1\over 4}\langle \overline{\psi},\psi\rangle \overline{\psi},
         \end{array} \right.
\end{equation}
where $\psi = \psi_1 - i\psi_2$. By ``odd'' we mean that the components $\psi_1$ and $\psi_2$ are anticommuting (Grassmann) variables.
An alternative elegant reformulation of this system can be obtained by adding the extra field $F:\Omega \longrightarrow \R^{n+1}$, which satisfies the 0th order PDE's
\begin{equation}\label{superharmonic0}
 F = {1\over 2i}\langle \psi,\overline{\psi}\rangle u
\end{equation}
and by setting
\[
 \Phi:= u + \theta^1\psi_1 + \theta^2\psi_2 + \theta^1\theta^2 F,
\]
where $\theta^1$ and $\theta^2$ are anticommuting coordinates, so that $(x,y,\theta^1,\theta^2)$ forms a complete system of coordinates on the {\em superplane} $\R^{2|2}$.
Then (\ref{superharmonic2+1}) and (\ref{superharmonic0}) are equivalent to
\begin{equation}\label{superharmonic}
 \overline{D}D\Phi + \langle \overline{D}\Phi,D\Phi\rangle \Phi = 0,
\end{equation}
where $D = {\partial \over \partial \theta} - \theta {\partial \over \partial z}$, $\overline{D} = {\partial \over \partial \overline{\theta}} - \overline{\theta} {\partial \over \partial \overline{z}}$.\\

\noindent
Actually, from (\ref{superharmonic2+1}) and (\ref{superharmonic0}) to (\ref{superharmonic}), we have used the fact that $u$, $\psi_1$, $\psi_2$ and $F$ are the components (supermultiplet) of a single map $\Phi$ from $\R^{2|2}$ to $S^n\subset \R^{n+1}$, which satisfies the superharmonic map equation (\ref{superharmonic}).\\

\noindent
Now we lift $\Phi$ to a framing supermap ${\cal F}: \R^{2|2}\longrightarrow SO(n+1)$ such that the composition of ${\cal F}$ with the projection $SO(n+1)\longrightarrow SO(n+1)/SO(n)\simeq S^n$ is $\Phi$. Set $\alpha:= {\cal F}^{-1}d{\cal F}$ and decompose $\alpha = \alpha_0 + \alpha_1$, according to the splitting of the Lie algebra $so(n+1)$ by the Cartan involution.\\

\noindent
Before giving a characterization of the superharmonic equation, it is useful to 
present a technical result concerning the exterior calculus of 1-forms on $\R^{2|2}$.
\begin{lemm} For a 1-form $\alpha$ on $\R^{2|2}$ with coefficients in a Lie algegra $\mathfrak{g}$, we have the equivalence
 \[
  d\alpha + {1\over 2}[\alpha\wedge \alpha] = 0
\quad \Longleftrightarrow \quad
\overline{D}\alpha(D) + D\alpha(\overline{D}) + [\alpha(\overline{D}), \alpha(D)] = 0.
 \]
\end{lemm}
{\em Remark:} $\Lambda^1(\R^{2|2})^*$ is spanned by $(d\theta, d\overline{\theta}, dz+(d\theta)\theta, d\overline{z} + (d\overline{\theta})\overline{\theta})$, the dual basis of $(D,\overline{D}, {\partial \over \partial z}, {\partial \over \partial \overline{z}})$. Hence in particular $\Lambda^2(\R^{2|2})^*$ is 6 dimensional. So the expansion of the l.h.s. of $d\alpha + {1\over 2}[\alpha\wedge \alpha] = 0$ leads to 6 equations which are a priori independant. The content of this lemma is that these 6 terms vanish as soon as one of these 6 coefficients (namely the coefficient of $d\theta\wedge d\overline{\theta}$) vanishes.\\

\noindent
Now the supermap ${\cal F}$ is superharmonic iff
\[
 \overline{D}\alpha_1(D) + [\alpha_0(\overline{D}), \alpha_1(D)] = 0.
\]
We hence deduce:
\begin{theo}
${\cal F}$ is superharmonic iff
\[
 \forall \lambda \in \C^*,\quad
\overline{D}\alpha(D)_\lambda + D\alpha(\overline{D})_\lambda + [\alpha(\overline{D})_\lambda, \alpha(D)_\lambda] = 0,
\]
where $\alpha(D)_\lambda:= \alpha_0(D) + \lambda^{-1}\alpha_1(D)$ and $\alpha(\overline{D})_\lambda:= \alpha_0(\overline{D}) + \lambda\alpha_1(\overline{D})$.
\end{theo}

\noindent
It results that this problem has the structure of a completely integrable system ({\sc F. O'Dea}, {\sc I. Khemar}). In particular the DPW algorithm for harmonic maps works.\\

\noindent
The DPW potential is a $\Lambda\mathfrak{g}^\C_\tau$-valued holomorphic 1-form $\mu$ on $\R^{2|2}$ s.t.
\[
 \mu(D) = \mu_0(D) + \theta\mu_\theta(D) = \lambda^{-1} (\cdot) + \lambda^0 (\cdot) + \cdots
\]
One integrates the equation
\[
 Dg = g\mu(D)
\]
 to get a holomorphic map $g = g_0 + \theta g_\theta: \R^{2|2}\longrightarrow \Lambda\mathfrak{G}^\C_\tau$. This implies in particular that
\[
 g_0^{-1}{\partial g_0\over \partial z} = - \left((\mu_0(D))^2 + \mu_\theta(D)\right) = \lambda^{-2}(\cdot) + \lambda^{-1} (\cdot) + \lambda^0 (\cdot) + \cdots
\]
Similarly, if ${\cal F} = {\cal F}_0 + \theta{\cal F}_\theta + \overline{\theta}{\cal F}_{\overline{\theta}} + \theta\overline{\theta}{\cal F}_{\theta \overline{\theta}}$, it turns out that ${\cal F}_0^{-1}d{\cal F}_0 = \lambda^{-2}(\cdot) + \lambda^{-1} (\cdot) + \lambda^0 (\cdot) + \lambda(\dot) + \lambda^2(\cdot)$. Hence we recover (for ${\cal F}_0$) something similar to a second order elliptic integrable system.

\subsection{Superprimitive maps \cite{k2}}
More precisely we can recover a second order elliptic integrable system close to the HSL surface theory in $\R^4$ by looking at {\em superprimitive maps} from $\R^{2|2}$ to the 4-symmetric space $SU(3)/SU(2)$: if $\Phi:\R^{2|2} \longrightarrow SU(3)/SU(2)$ is a superprimitive map then the first component $u$ in the decomposition $\Phi = u + \theta^1\psi_1 + \theta^2\psi_2 + \theta^1\theta^2F$ is a conformal HSL immersion (with the restriction that the Lagrangian angle $\beta$ is equal to a {\em real} constant plus a harmonic non constant {\em nilpotent} function).

\end{document}